\documentclass[preprint,3p,times]{elsarticle}

\usepackage{amssymb}
\usepackage{indentfirst}
\usepackage{subfigure}
\usepackage{textcomp}
\usepackage{graphicx}
\usepackage{amsmath}
\usepackage{amsopn}
\usepackage{amssymb}
\usepackage{url}
\usepackage{enumerate}

\journal{arXiv}

\begin{document}

\begin{frontmatter}



\title{Quantum Strategies Win in a Defector-Dominated Population}


\author{Qiang Li$^{\textrm{a,b}}$, Azhar Iqbal$^{\textrm{b}}$, Minyou Chen$^{\textrm{a}}$, Derek Abbott$^{\textrm{b}}$}

\address{$^{\textrm{a}}$ College of Electrical Engineering, Chongqing University, Chongqing 400030, China\\
$^{\textrm{b}}$ School of Electrical and Electronic Engineering,
University of Adelaide, SA 5005, Australia}

\begin{abstract}
Quantum strategies are introduced into evolutionary games. The
agents using quantum strategies are regarded as invaders whose
fraction generally is 1\% of a population in contrast to the 50\%
defectors. In this paper, the evolution of strategies on networks is
investigated in a defector-dominated population, when three networks
(Regular Lattice, Newman-Watts small world network, scale-free
network) are constructed and three games (Prisoners' Dilemma,
Snowdrift, Stag-Hunt) are employed. As far as these three games are
concerned,
the results show that quantum strategies can always invade the
population successfully. Comparing the three networks, we find that
the regular lattice is most easily invaded by agents that adopt
quantum strategies. However, for a scale-free network it can be
invaded by agents adopting quantum strategies only if a hub is
occupied by an agent with a quantum strategy or if the fraction of
agents with quantum strategies in the population is significant.
\end{abstract}

\begin{keyword}
Quantum Computation\sep Quantum Game\sep Multi-Agent System\sep
Evolutionary Game\sep Strategy Evolution

\end{keyword}

\end{frontmatter}


\section{Introduction}
Game theory has been widely applied in both social and biological
fields, in order to describe interactions between agents. Recently,
the evolution of behavior of agents in a population, in the
framework of evolutionary games on graphs, has attracted much
interdisciplinary attention. Nowak and
May~\cite{Nowak1992,Nowak1993} firstly introduced the spatial
Prisoner's Dilemma (PD) game, in which agents (players) occupy all
vertices of a two-dimensional lattice and the edges represent
neighbor relations between the corresponding agents. This pioneering
work triggered an intensive investigation of spatial games and the
PD game is a model frequently adopted by
researchers~\cite{Szolnoki2009,Szabo2009}. It is known that the
structure of the network is also a key factor in the evolution of
behavior of agents. Later, a shift from evolutionary games on
regular lattices to evolutionary games on complex networks was
proposed~\cite{Perc2010}, in particular on small world
networks~\cite{Shang2006,Wu2007,Chen2008} and on scale-free
networks~\cite{Assenza2008,Lee2008,Perc2009}. Meanwhile, other
games, such as Snowdrift (SD)~\cite{Taylor1978}, Stag-Hunt
(SH)~\cite{Skyrms2004}, and Public Goods games, have produced
interesting
results~\cite{Hauert2004,Szolnoki2009a,Helbing2010,Starnini2011}.

Surprisingly, the concept of evolutionary games has been extended to
the microworld to describe interactions of biological
molecules~\cite{Turner1999,Pfeiffer2001,Frick2003,Chettaoui2007}, a
domain where quantum mechanics defines the laws. Meanwhile, game
theory is also generalized to the quantum regime, and a new area
called quantum game theory has emerged from the field of quantum
computation. In recent years, much interest has been focused on
quantum game theory. For instance, Meyer's results~\cite{Meyer1999}
showed that if an agent in a penny flip game is allowed to implement
quantum strategies, she/he can always defeat her/his opponent
playing a classical strategy and can thus increase her/his expected
payoff. Eisert et al.~\cite{Eisert1999} quantized the PD and
demonstrated that it is possible to escape the dilemma when both
players resort to quantum strategies. Marinatto et
al.~\cite{Marinatto2000} found a unique equilibrium for the Battle
of the Sexes game, when entangled strategies were allowed. Later,
evolutionarily stable strategies in quantum games and an
evolutionary quantum game were also studied by Iqbal et
al.~\cite{Iqbal2001a} and Kay et al.~\cite{Kay2001} respectively.
Moreover, quantum games have also been implemented using quantum
computers~\cite{Du2002,Prevedel2007,Schmid2009}. For further
background on quantum games, see~\cite{Flitney2002,Guo2008}.

In quantum game theory, agents are allowed to use quantum strategies
from a quantum strategy set that is a much larger set than a
classical one, i.e., a classical strategy set is only a subset of a
quantum strategy set. This larger space offers a possibility for a
diversity of agent behavior and allows new patterns to emerge. In
this paper, we assume all agents in a population are quantum agents
who can use quantum strategies to play games with their neighbors
and make decisions. However, initially, only a few randomly selected
agents, about 1\% in the population, are assigned quantum
strategies, while the others are players with strategies taken from
the classical strategy set. The fraction of defectors is about half
of the population. This work discusses how quantum strategies spread
in the population and how strategies evolve over repeated games
played on networks. Therefore, three networks (Regular Lattice,
Newman-Watts small world network, scale-free network) are
constructed and three games (Prisoners' Dilemma, Snowdrift,
Stag-Hunt) are employed. The games encapsulate agents' responses to
different external stimuli, while those networks provide different
environments for agents. It is worth noting that a quantum strategy
is not a probabilistic sum of pure classical strategies (except
under special conditions) and it also cannot be reduced to the pure
classical strategies~\cite{Iqbal2001a}.

The rest of this paper is organized as follows: Section 2 briefly
introduces some concepts of quantum computation and quantum games.
Next, the model and the simulation setup is described in Section 3
and Section 4 respectively. In Section 5, results are demonstrated
firstly. Later, the situation of strategies spreading on networks is
discussed and the evolution of strategies is analyzed when different
games are adopted. The conclusion is given in Section 6.

\section{Quantum Games}
Before introducing quantum games, we describe some basic concepts of
quantum computation. In quantum computation, a qubit is the
elementary unit, which is typically a microscopic system, such as a
nuclear spin or a polarized photon, while the Boolean states 0 and 1
are represented by a prescribed pair of normalized and mutually
orthogonal quantum states labeled as $\{|0\rangle,|1\rangle\}$ to
form a `computational basis'~\cite{Ekert2001}. Furthermore, any pure
state of the qubit can be written as a superposition state
$\alpha|0\rangle+\beta|1\rangle$ for some $\alpha$ and $\beta$
satisfying $|\alpha|^{2}+|\beta|^{2}=1$~\cite{Ekert2001}. Also, if
any manipulations on qubits are needed, they have to be performed by
unitary operations, which can be carried out by a quantum logic gate
or a quantum circuit~\cite{Ekert2001}. The most often used quantum
gate is the Hadamard gate. If a qubit in state $|0\rangle$ or
$|1\rangle$ is manipulated by it, the qubit will be in the following
state
\begin{equation}
\left\{ \begin{array}{cc} |0\rangle\xrightarrow{\hat{H}}\frac{1}{\sqrt{2}}|0\rangle+\frac{1}{\sqrt{2}}|1\rangle \vspace{4pt}\\
|1\rangle\xrightarrow{\hat{H}}\frac{1}{\sqrt{2}}|0\rangle-\frac{1}{\sqrt{2}}|1\rangle\end{array}
\right., \hat{H}=\frac{1}{\sqrt{2}} \left( \begin{array}{cc}
1 & 1\\
1 & -1
\end{array} \right).
\end{equation}
For further details, see~\cite{Ekert2001,Nielsen2000}.

In the following, we take the PD as an example for introducing
quantum games. As is known, PD can be used to model many strategic
phenomena in the real world and it has been widely applied in a
number of scientific fields. In this symmetric game, each of two
players has two available strategies, Cooperation (C) and Defection
(D). Next, each of the two players chooses a strategy against the
other's at the same time, but both sides do not know the opponent's
strategy. Finally, each agent acquires a payoff, where the payoff
matrix to the first agent can be written as
\begin{equation}
\bordermatrix{%
  & C & D\cr
C & R & S\cr D & T & P\cr }=
\bordermatrix{%
  & C & D\cr
C & 1 & 0\cr D & 2 & 0\cr }.
\end{equation}
According to the conclusion in classical game theory, the strategy
profile $(D,D)$ is the unique Nash Equilibrium
(NE)~\cite{Nash1950,Nash1951}, but unfortunately the strategy
profile $(C,C)$ is merely the best choice that is Pareto
optimal~\cite{Fudenberg1983}. Hence, the dilemma is produced.

\begin{figure}[htbp]
\centering
\includegraphics[width=0.5\textwidth]{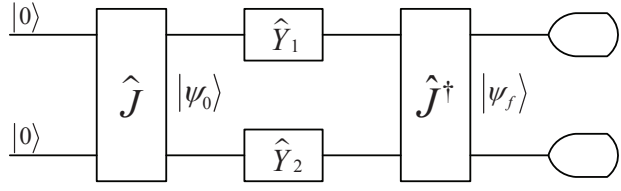}
\caption{The block diagram for the Eisert's scheme.}\label{fig:1}
\end{figure}

However, Eisert et al. quantized the PD game and introduced an
elegant scheme, the physical model of a quantum game, which is shown
in Fig.~\ref{fig:1}~\cite{Eisert1999}. According to their results,
the dilemma in the classical counterpart can be escaped in a
restricted strategic space~\cite{Eisert1999}, when quantum
strategies are used.
In their model, at first two basis vectors
$\{|C=0\rangle,|D=1\rangle\}$ in Hilbert space are assigned to the
possible outcomes of the classical strategies, $C=0$ and $D=1$
respectively~\cite{Eisert1999}. Then, suppose the initial state is
$|\psi_{0}\rangle=\hat{J}~|00\rangle$ before the game is played,
where $\hat{J}$ is an entangling operator that is known to both
players. For a $2\times2$ game, the entangling operator $\hat{J}$
has form below~\cite{Benjamin2001,Du2001}
\begin{equation}
\hat{J}(\omega)=\exp(i\frac{\omega}{2}\sigma_{x}^{\otimes2})=I^{\otimes2}\cos\frac{\omega}{2}+i\sigma_{x}^{\otimes2}\sin\frac{\omega}{2}
\end{equation}
where $\omega\in[0,\pi/2]$ is a measure of entanglement of a game.
When $\omega=\pi/2$, there is a maximally entangled game, in which
the entangling operator can be written as
\begin{equation}
\hat{J}(\omega)=\frac{1}{\sqrt{2}}~(I^{\otimes2}+i\sigma_{x}^{\otimes2}).
\end{equation}
Next, each agent chooses a unitary operator $\hat{Y}$ as a strategy
from the two-parameter strategy space
$\hat{S}$~\cite{Eisert1999}
\begin{equation}
\hat{Y}(\gamma,\phi)=\begin{pmatrix}e^{i\phi}\cos\frac{\gamma}{2} &
\sin\frac{\gamma}{2} \\ -\sin\frac{\gamma}{2} &
e^{-i\phi}\cos\frac{\gamma}{2}\end{pmatrix}\in \hat{S},
\end{equation}
where $\gamma\in[0,\pi]$, $\phi\in[0,\pi/2]$. Then, she/he operates
it on the qubit that belongs to her/him, which makes the game in a
state $(\hat{Y_{1}}\otimes\hat{Y_{2}})\hat{J}|00\rangle$. In the
end, before a projective measurement on the basis
$\{|0\rangle,|1\rangle\}$ is carried out, the final state is
\begin{equation}
|\psi_{f}\rangle=\hat{J}^{\dag}(\hat{Y_{1}}\otimes\hat{Y_{2}})\hat{J}~|00\rangle.
\end{equation}
As such, the first agent's expected payoff is written as
\begin{equation}\label{Eq2}
\Pi(\gamma,\phi)=R|\langle\psi_{f}|00\rangle|^{2}+S|\langle\psi_{f}|01\rangle|^{2}+T|\langle\psi_{f}|10\rangle|^{2}+P|\langle\psi_{f}|11\rangle|^{2}.
\end{equation}

\section{The Model}
Assume there is an undirected network $G(V,E)$ with $N$ nodes, in
which $V$ is the set of nodes and $E$ is the set of links. Also,
each node $i\in V$ is occupied by an agent and its neighbor $j$ is
any other agent such that there is a link between them, so the set
of neighbors of an agent $i$ can be defined as
$\Gamma_{i}=\{j|e_{ij}\in E, j\in V\backslash i\}$.

In this paper, three different networks will be constructed. They
are a Regular Lattice (RL) with periodic boundary conditions, a
Newman-Watts (NW) small world network~\cite{Newman1999,Newman1999a}
and a Scale-Free (SF) network~\cite{Barabasi1999,Albert2002}. When
periodic boundary conditions are involved, they can guarantee each
node in the regular lattice has four neighbors. In addition, for
avoiding isolated nodes, the NW small world network is selected in
our work instead of the Watts-Strogatz (WS) network. The NW network
can be established in two steps~\cite{Newman2003}. At first, a
regular lattice, with periodic boundary conditions, is constructed,
and then links are added with probability $p_{\rm nw}$ between any
two randomly chosen nodes. Finally, the SF network is established
according to the Barab\'{a}si-Albert model~\cite{Barabasi1999} whose
algorithm consists of two steps, growth and preferential attachment.
It can start with a small network of $m_{0}\ll N$ all connected
nodes, and then a new node with $m\leq m_{0}$ links will be added to
the network. Its $m$ links will be connected to $m$ different nodes
chosen with probability $p_{\rm sf}(i)$ which can be calculated as
below,
\begin{equation}
p_{\rm sf}(i)=\frac{k_{i}}{\sum_{j\in V}k_{j}}.
\end{equation}
Here, $k$ is the degree of a node. This procedure will be repeated
many times till the number of nodes of the network is $N$.

Initially, each agent on the network is assigned a strategy randomly
from the set of strategies $\{\hat{C},\hat{D},\hat{H},\hat{Q}\}$.
Next, an agent $i$ will play a $2\times2$ entangled quantum game in
turn with each one of its $|\Gamma_{i}|$ neighbors according to the
physical model of a quantum game (Fig.~\ref{fig:1}), where the
symbol $|\cdot|$ is the cardinality of a set. Throughout the paper,
all quantum games are maximally entangled games, if not otherwise
explicitly stated. And then its expected payoff
$\Pi_{ij},j\in\Gamma_{i}$ can be calculated by Eq.~\ref{Eq2}. The
agent's total payoff $F_{i}$ is obtained by accumulating all it
receives $F_{i}=\sum_{j\in\Gamma_{i}}\Pi_{ij}$.

After that, it will choose a neighbor from its neighborhood randomly
and imitate its strategy with probability $p_{i}$~\cite{Santos2005},
\begin{equation}
p_{i}=\bigg\{\begin{array}{lll}
\frac{F_{j}-F_{i}}{\alpha\cdot\textrm{max}(k_{i},k_{j})}, & \textrm{ $F_{j}>F_{i}, j\in\Gamma_{i}$}\vspace{4pt}\\
0, & \textrm{ otherwise}
\end{array},
\end{equation}
where $\alpha$ is a constant that can be calculated as below
according to different games~\cite{Szabo2007},
\begin{equation}
\alpha=\left\{
\begin{array}{ll}
    T-S, & \hbox{PD} \vspace{4pt}\\
    T-P, & \hbox{SD} \vspace{4pt}\\
    R-S, & \hbox{SH} \hspace{4pt}\textrm{.}
  \end{array}
\right.
\end{equation}
After all the agents acquire their payoffs, their strategies are
updated synchronously. This process will be repeated by a maximum
number of $10^{4}$ generations and the fractions of agents with
different strategies are obtained by averaging the last 1000
generations, which produces a result of evolution of strategies. The
final result is obtained by averaging over at least 100 of these
results. If strategies of all agents do not change for 500
consecutive generations, it is deemed that an equilibrium has been
reached and the iterations are stopped.

\section{Simulation setup}
Assume a population of $50\times50$ agents are located at nodes of
the above mentioned networks. For the NW network, the probability
that links are added between any two randomly chosen nodes is
$p_{\rm nw}=0.5$, while for the SF network, the number of nodes of
the initial core network is $m_{0}=2$ or $3$ and the links of each
new node are set at $m=2$. Throughout all simulations, the network
topology remains static. In this paper, we consider two sets of
strategies:
\begin{enumerate}[{Case} 1.]
\item There are three strategies in the set $\hat{S_{1}}=\{\hat{C},\hat{D},\hat{H}\}$. Initially, three strategies, $\hat{C}$ (Cooperation), $\hat{D}$ (Defection), $\hat{H}$ (Hadamard), are assigned to agents randomly and the
fractions of strategies are 49\%, 50\% and 1\% respectively. Here,
the unitary
operators $\hat{C}$ and $\hat{D}$ have the forms below
\begin{equation}
\hat{C}=\begin{pmatrix}1 & 0 \\ 0 &
1\end{pmatrix},\hspace{4pt}\hat{D}=\begin{pmatrix}0 & 1
\\ 1 & 0\end{pmatrix}.
\end{equation}
\item There are four strategies in the set $\hat{S_{2}}=\{\hat{C},\hat{D},\hat{H},\hat{Q}\}$. Initially, four strategies, $\hat{C}$, $\hat{D}$, $\hat{H}$ and $\hat{Q}$, are assigned to the population randomly and the
fractions of strategies are 49\%, 49\%, 1\% and 1\% respectively.
The strategy $\hat{Q}$ takes the form
\begin{equation}
\hat{Q}=\begin{pmatrix}i & 0 \\ 0 &
-i\end{pmatrix}.
\end{equation}
\end{enumerate}
In these two cases, the quantum strategy $\hat{H}$ brings a
\textit{miracle move}~\cite{Eisert1999} when an agent uses it
against the other's classical strategy. Also, the quantum strategy
profile $(\hat{Q},\hat{Q})$ is a new NE observed by Eisert et
al.~\cite{Eisert1999}, when players choose their strategies from the
strategy space $\hat{S}=\hat{Y}(\gamma,\phi)$.%

Then, the PD, SD and SH games are played by all agents on RL, NW,
and SF networks respectively. To be compatible with previous studies
and without loss of generality, the payoff matrix of the PD game is
chosen as $R = 1$ (Reward), $T = b$ ($1<b\leq2$) (Temptation), $P =
0$ (Punishment) and $S = 0$ (Sucker's payoff) satisfying the
inequalities $T>R>P>S$; the payoff matrix of the SD game as $T =
b>1$, $R = b-c/2$, $S = b-c$ and $P = 0$ satisfying $T
>R >S>P$, and the cost-to-benefit ratio of mutual cooperation is
defined as $r = c /(2b - c)$~\cite{Hauert2004}, where $c = 1$ and
$0<r\leq1$; the payoff matrix of the SH game as $R = 1$, $T = r$
($0<r\leq1$), $S = -r$ and $P = 0$ satisfying $R>T>P>S$.

The payoffs under any strategy profiles can be calculated according
to the model of a quantum game in Section 2. As is known, the
strategy profiles $(\hat{H},\hat{H})$ and $(\hat{Q},\hat{Q})$ are
Nash equilibria for PD, SD and SH games in Case 1 and Case 2
respectively. However, it is worth noting that for the SH game there
are two Nash equilibria, i.e., besides $(\hat{H},\hat{H})$ or
$(\hat{Q},\hat{Q})$, the other is $(\hat{C},\hat{C})$.

\section{Results and Discussion}
Throughout all simulations, the agents with quantum strategies
$\hat{H}$ and $\hat{Q}$ are regarded as invaders, so their fractions
are restricted to 1\% of the whole population initially. Thus, our
simulations focus on investigating how the quantum strategies invade
the population and how strategies evolve on networks. In order to
elucidate the affects of quantum strategies, three games are
considered to observe the behavior of quantum strategies.
Furthermore, the structure of a network is also a key factor for the
evolution of strategies, so three different networks (RL, NW and SF)
are constructed to test their influences.

\begin{figure}[htbp]
\centering \subfigure[]{
\includegraphics[width=0.32\textwidth]{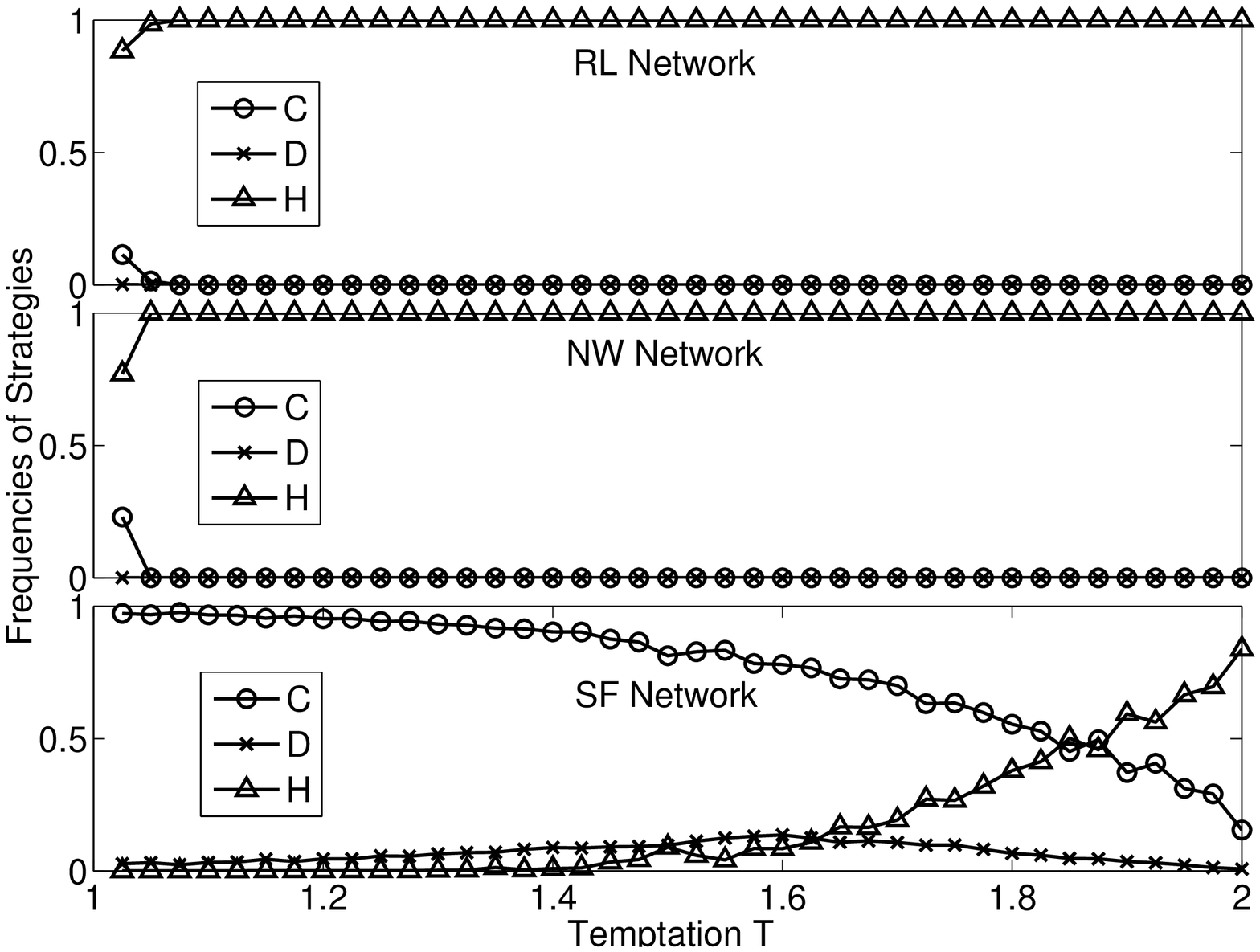}} \subfigure[]{
\includegraphics[width=0.32\textwidth]{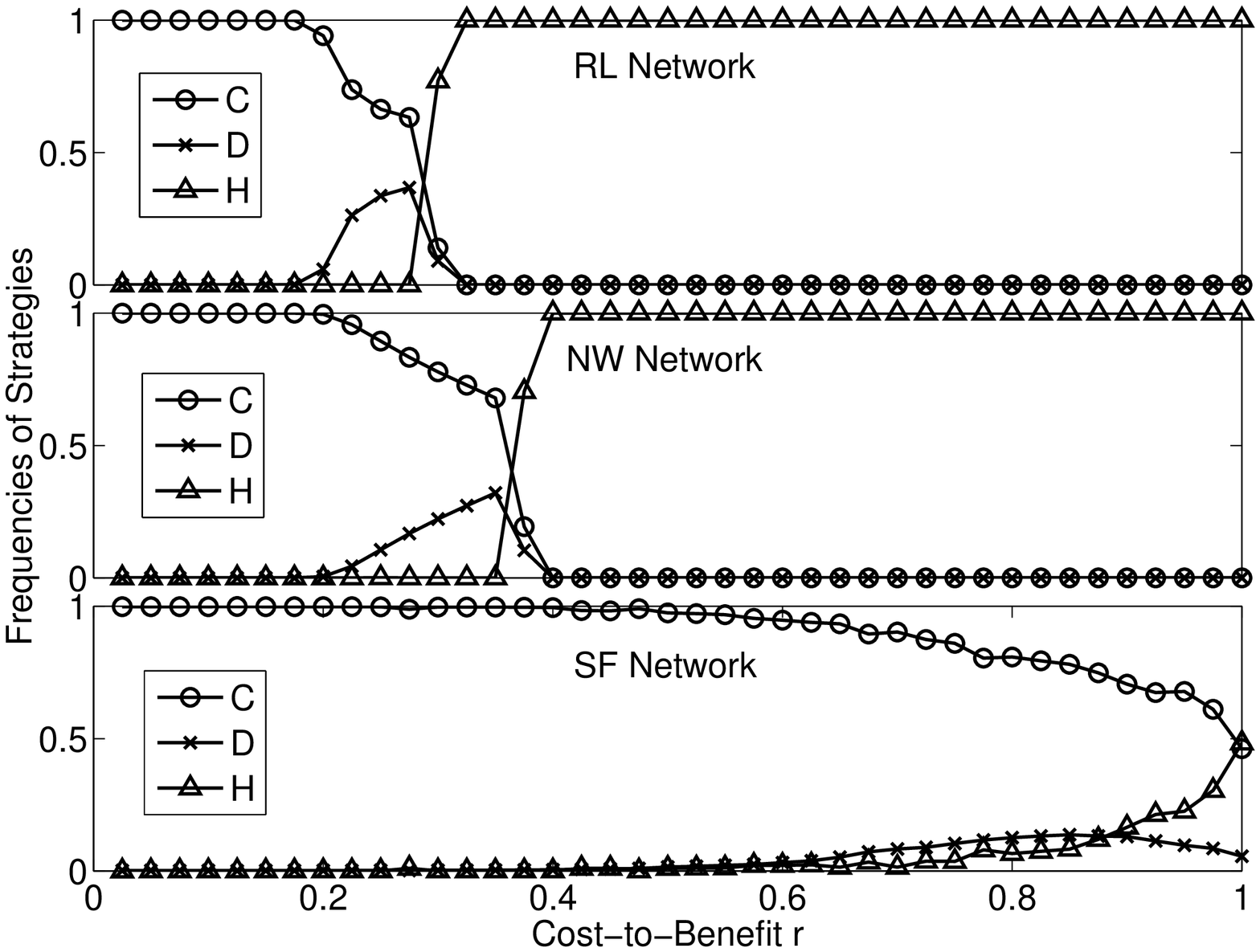}}
\subfigure[]{ \includegraphics[width=0.32\textwidth]{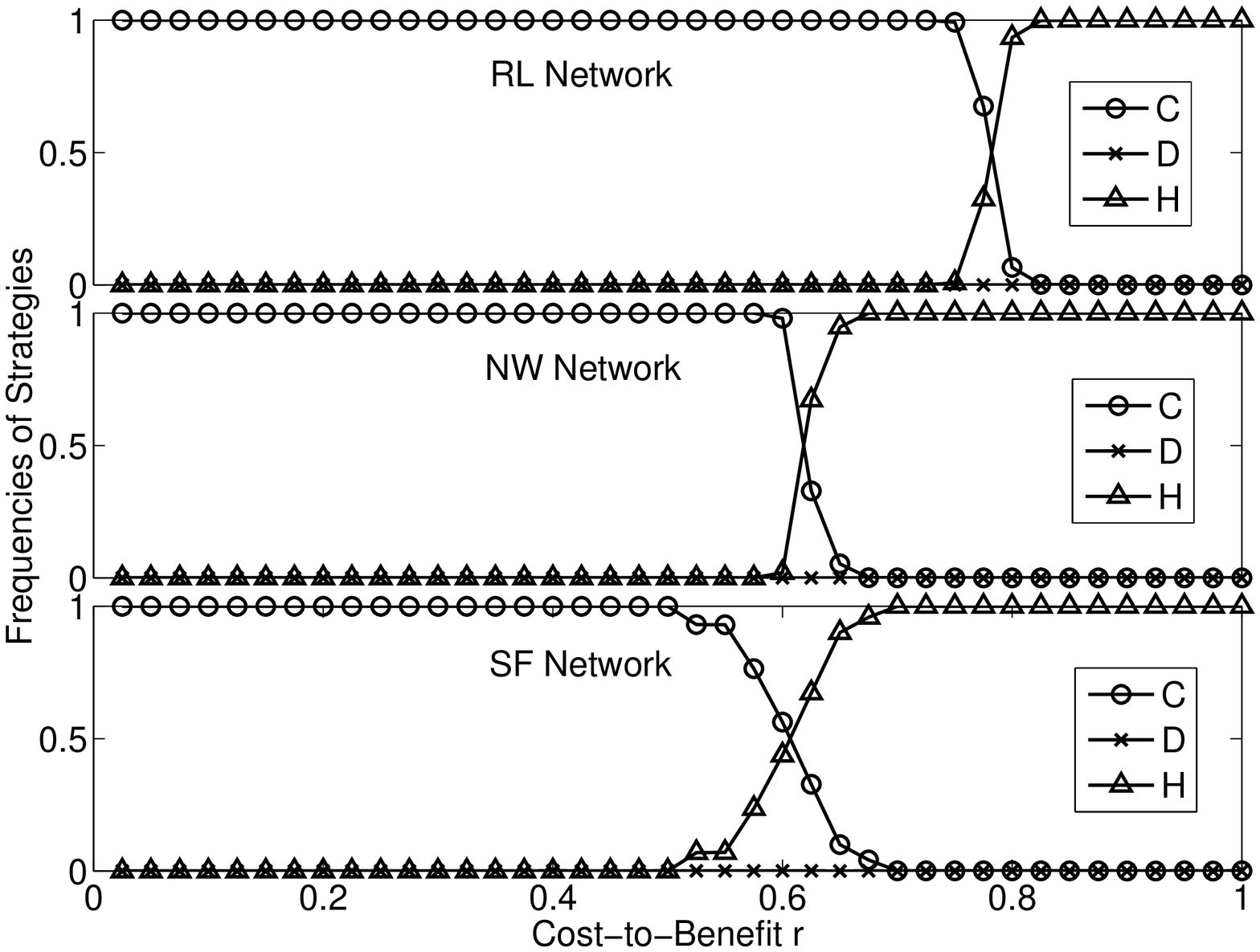}}\\
\caption{Fractions of agents with three strategies in Case 1 on
different games and networks. (a) Prisoners' Dilemma. (b) Snowdrift.
(c) Stag-Hunt. In the upper, median and lower sub-figures of (a),(b)
and (c), games are played on RL, NW and SF networks
respectively.}\label{fig1}
\end{figure}

For Case 1, the upper and median sub-figures of Fig.~\ref{fig1} show
that the fractions of agents with three strategies on RL and NW
networks are similar. Given our three games, the quantum strategy
$\hat{H}$ becomes the Evolutionarily Stable Strategy (ESS) almost
from the outset when the PD game is played, i.e., even if the
temptation $T$ is only a little larger than the reward $R$, the
strategy $\hat{H}$ can dominate the network successfully. However,
for SD and SH games, it is more difficult for the strategy $\hat{H}$
to be the ESS. When the SD game is employed, the strategy $\hat{H}$
is played by all agents on the network, only if the cost-to-benefit
ratio $r\geq0.32$ and $r\geq0.40$ on RL and NW networks
respectively. Before this value, almost no agents use the strategy
$\hat{H}$, while at that time the frequencies of the strategies
$\hat{C}$ and $\hat{D}$ are similar to the case of a SD game without
a quantum strategy. For the SH game, a coordination game, the
strategy $\hat{H}$ invades the network successfully when $r\geq0.82$
and $r\geq0.67$ respectively on RL and NW networks. From above
analysis, it can be said that the SH game and the structure of NW
network is more advantageous for classical strategies than the
quantum ones.

\begin{figure}[th]
\centering \subfigure[]{
\includegraphics[width=0.32\textwidth]{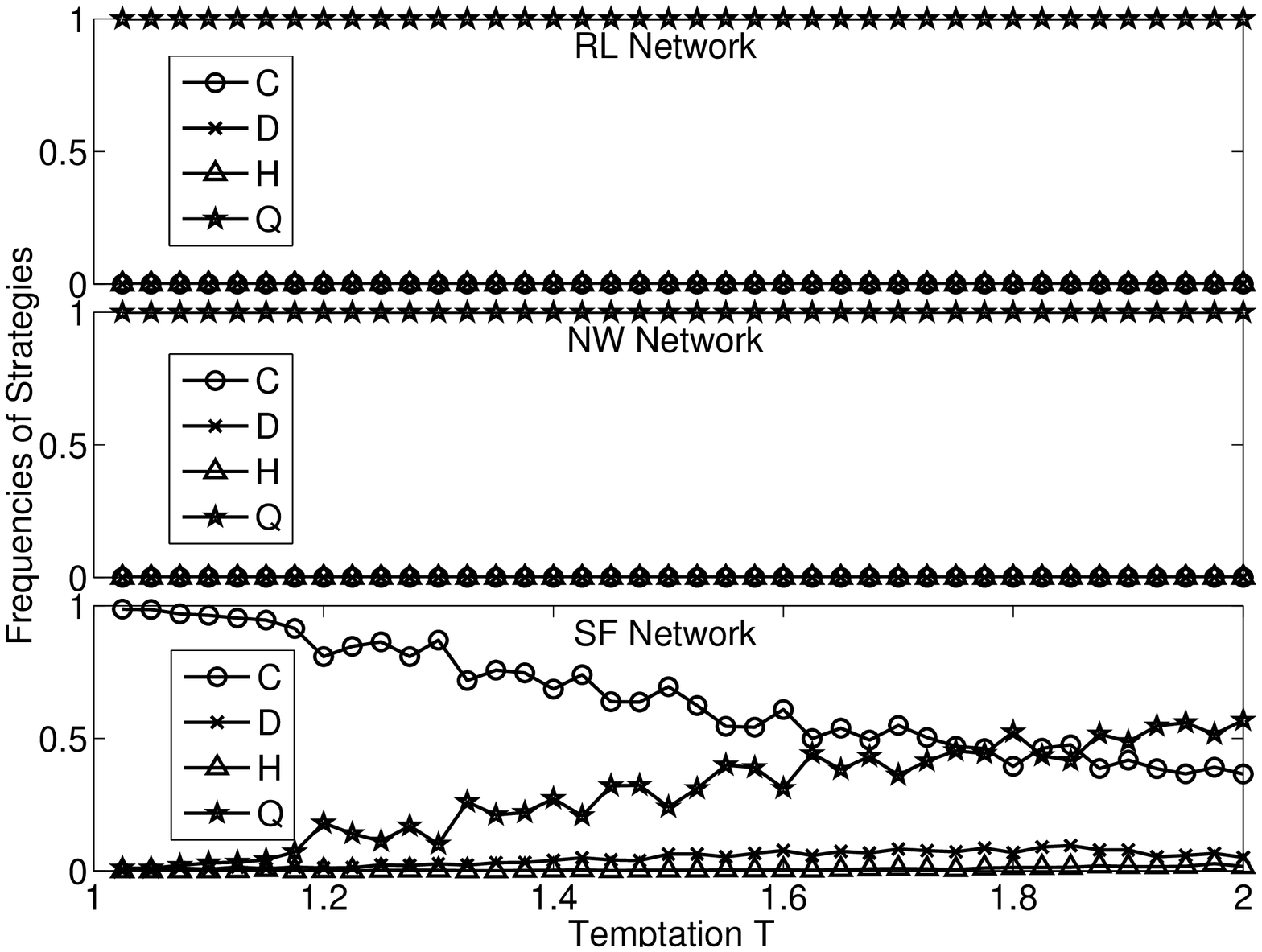}} \subfigure[]{
\includegraphics[width=0.32\textwidth]{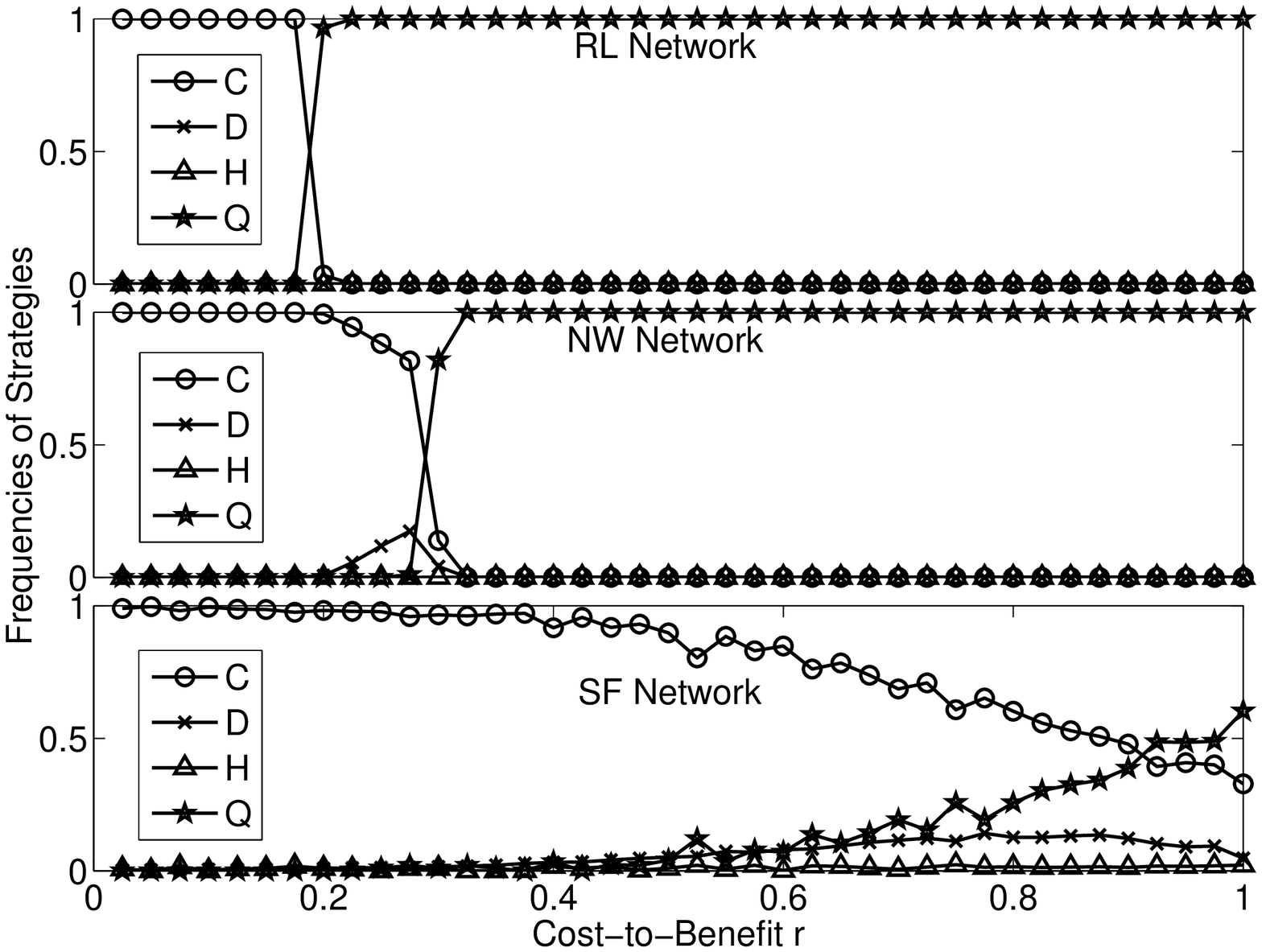}}
\subfigure[]{ \includegraphics[width=0.32\textwidth]{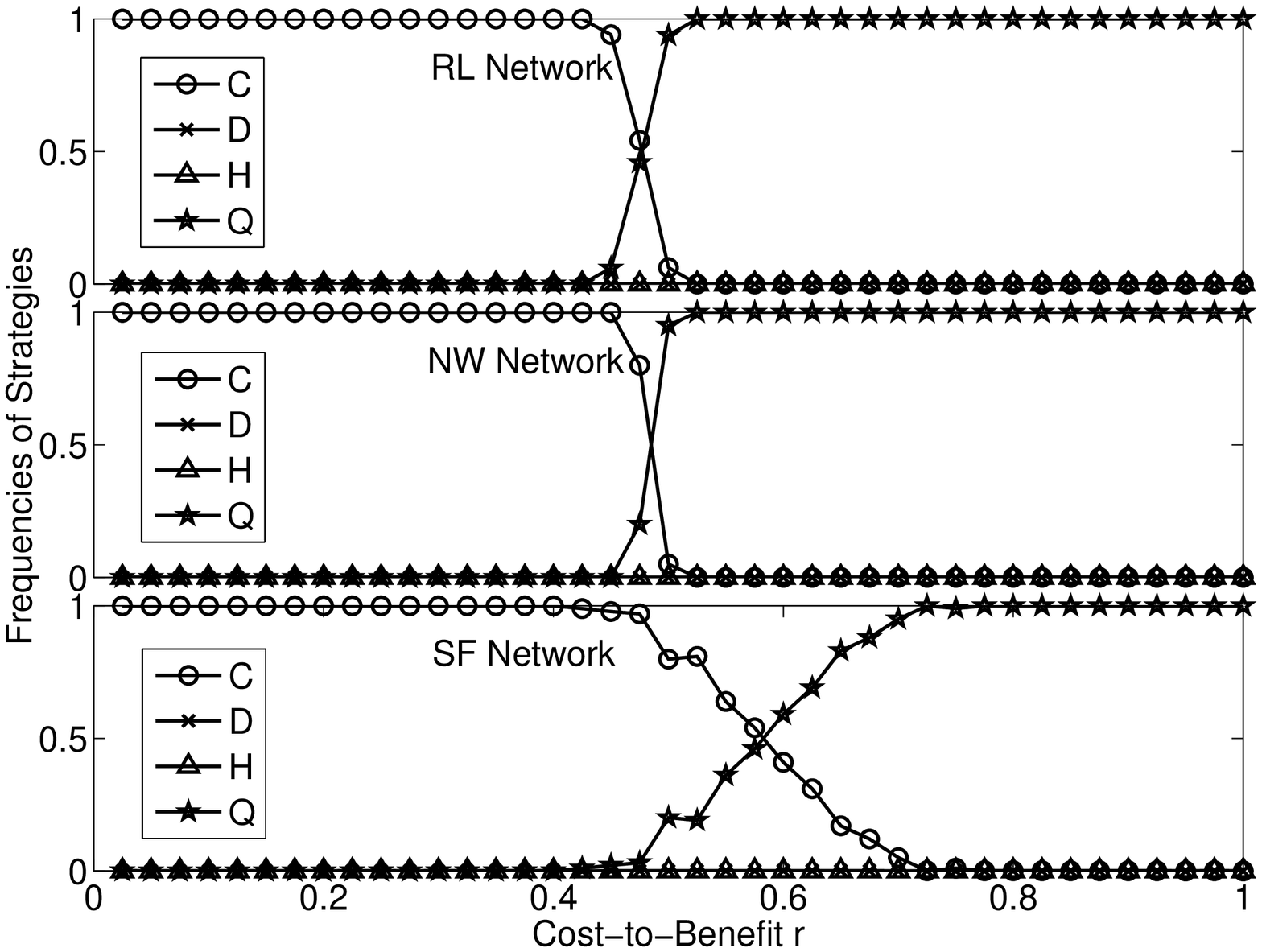}}\\
\caption{Fractions of agents with four strategies in Case 2 on
different games and networks. (a) Prisoners' Dilemma. (b) Snowdrift.
(c) Stag-Hunt. In the upper, median and lower sub-figures of (a),(b)
and (c), games are played on RL, NW and SF networks
respectively.}\label{fig2}
\end{figure}

The results of evolution of strategies in Case 2 are shown in
Fig.~\ref{fig2}. From Fig.~\ref{fig2}, it can be seen that for the
PD the strategy $\hat{Q}$ can invade the whole network from the
outset on both the RL and NW networks, and the strategy $\hat{Q}$
becomes the dominate strategy over $\hat{H}$ in Case 1 when the SD
and SH games are adopted, which is improved about 10\%,
8\% (SD on RL and NW) 
 and 30\%, 15\% (SH on RL and NW) 
  respectively. In contrast to Case 1,
it can be inferred that the strategy $\hat{Q}$ is more aggressive
than the strategy $\hat{H}$.

\begin{figure}[th]
\centering \subfigure[]{
\includegraphics[width=0.32\textwidth]{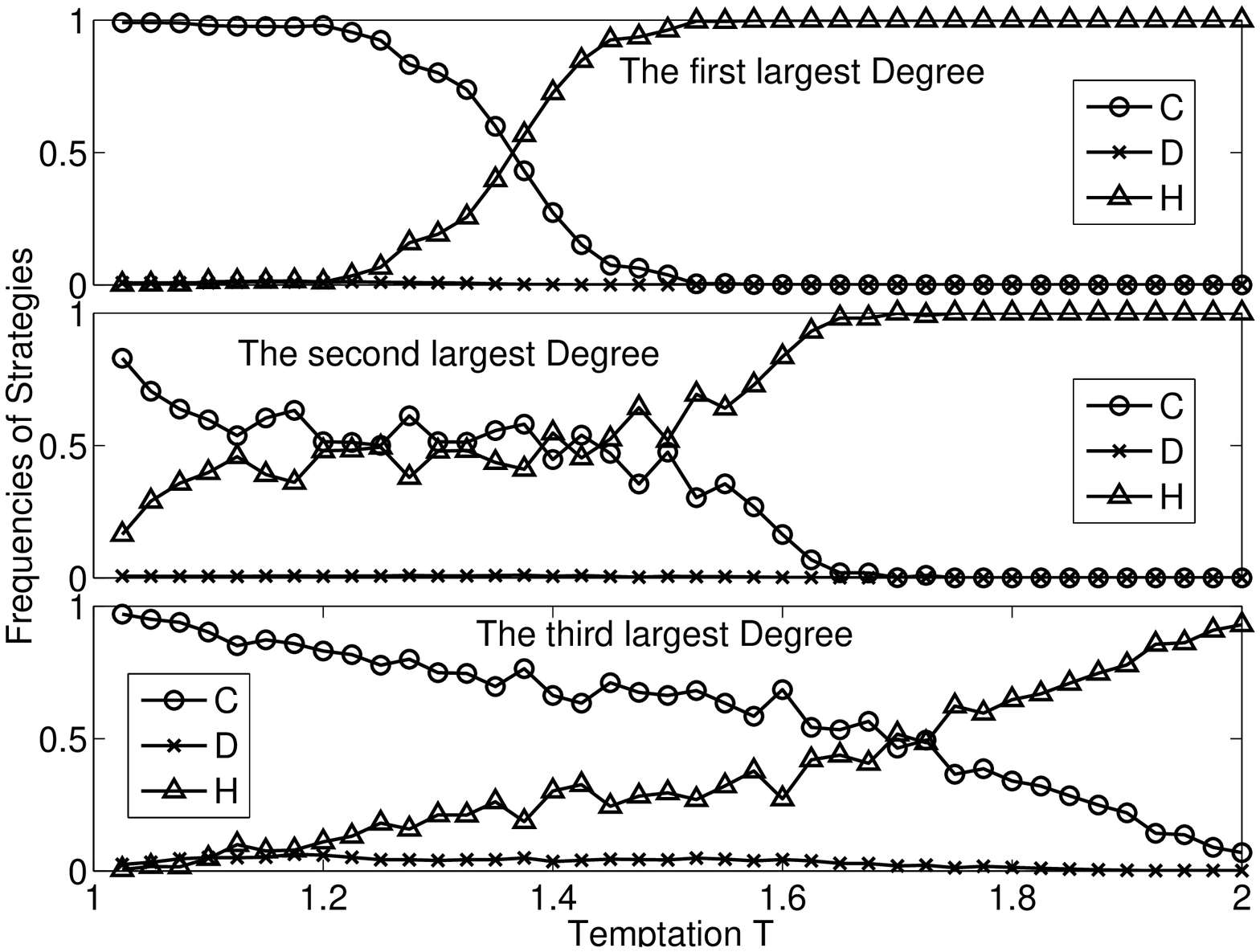}} \subfigure[]{
\includegraphics[width=0.32\textwidth]{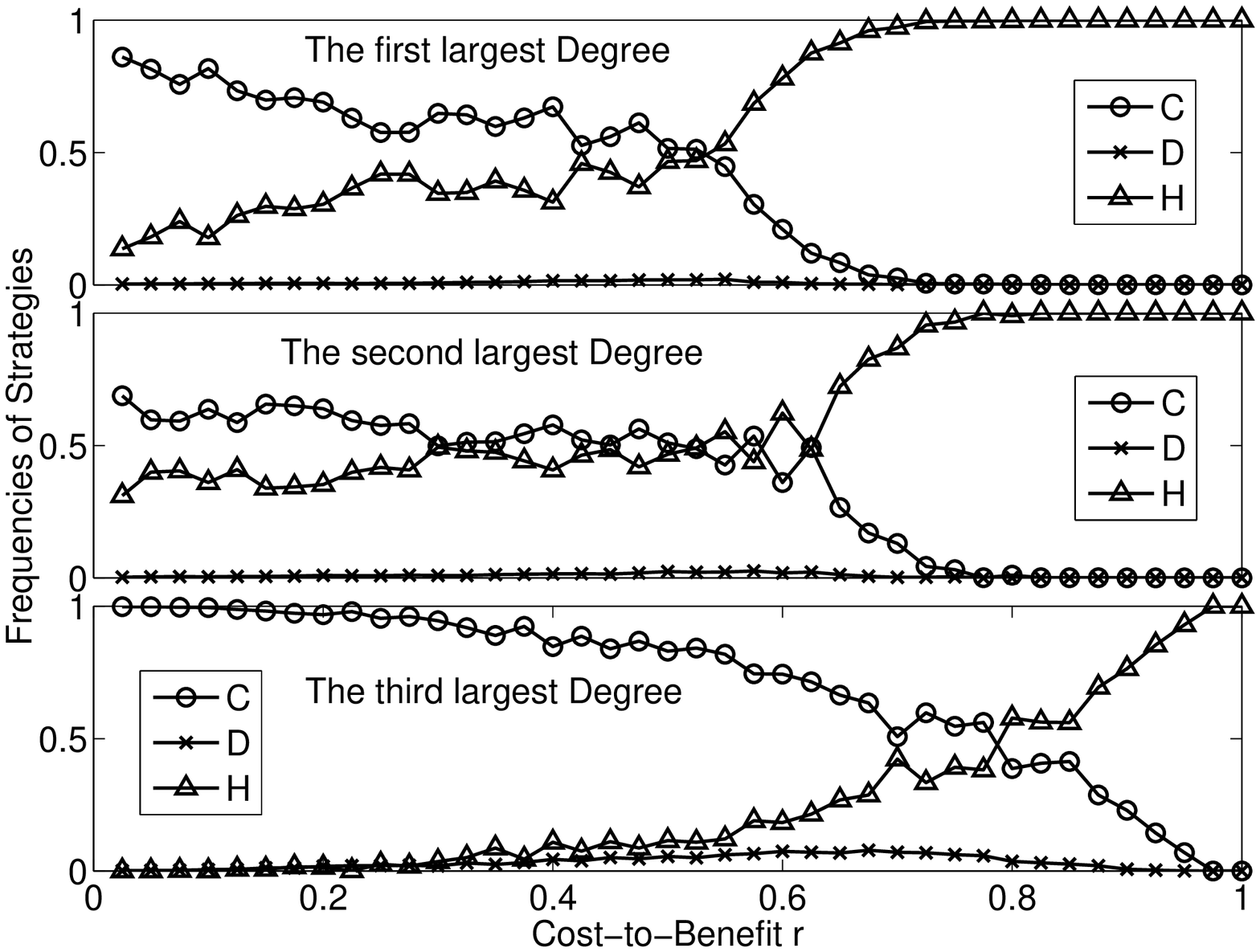}}
\subfigure[]{ \includegraphics[width=0.32\textwidth]{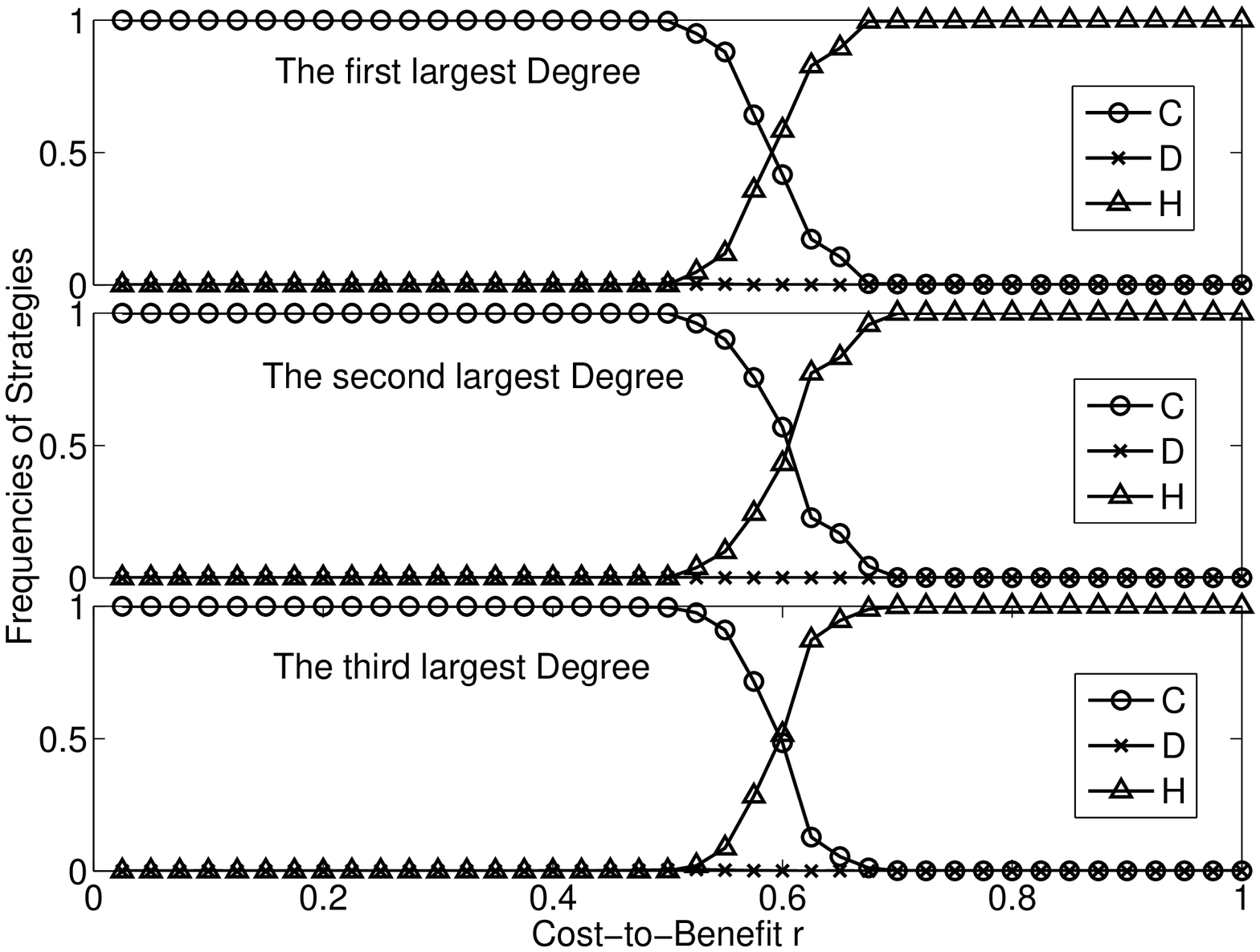}}\\
\caption{Fractions of agents with different strategies in Case 1 on
a SF network, when the agents with the first three largest degrees
play the quantum strategy $\hat{H}$. (a) Prisoners' Dilemma. (b)
Snowdrift. (c) Stag-Hunt. In the upper, median and lower sub-figures
of (a),(b) and (c), the agent with the first, second or third degree
is assigned the quantum strategy $\hat{H}$
respectively.}\label{fig3}
\end{figure}

\begin{figure}[th]
\centering \subfigure[]{
\includegraphics[width=0.32\textwidth]{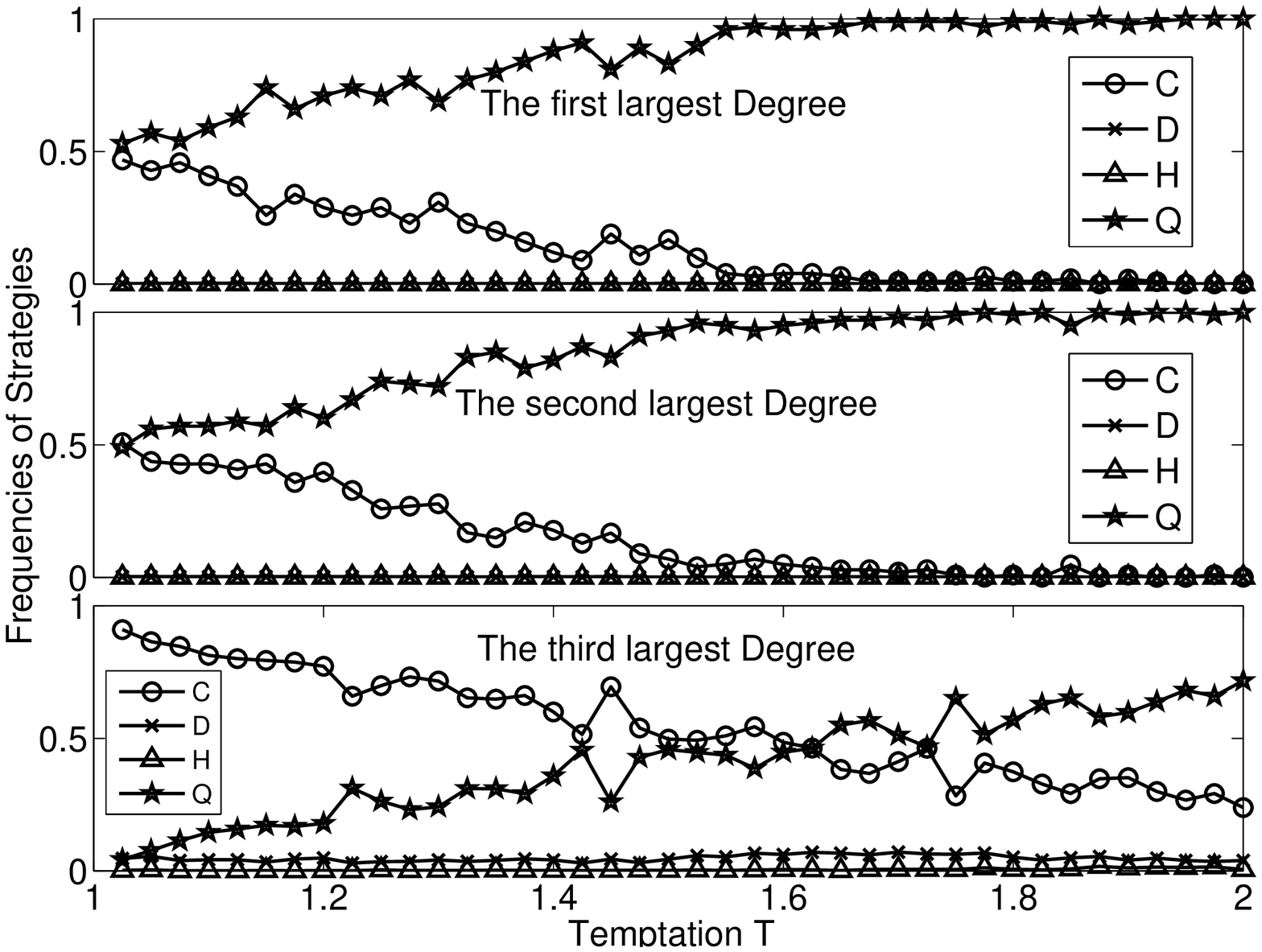}} \subfigure[]{
\includegraphics[width=0.32\textwidth]{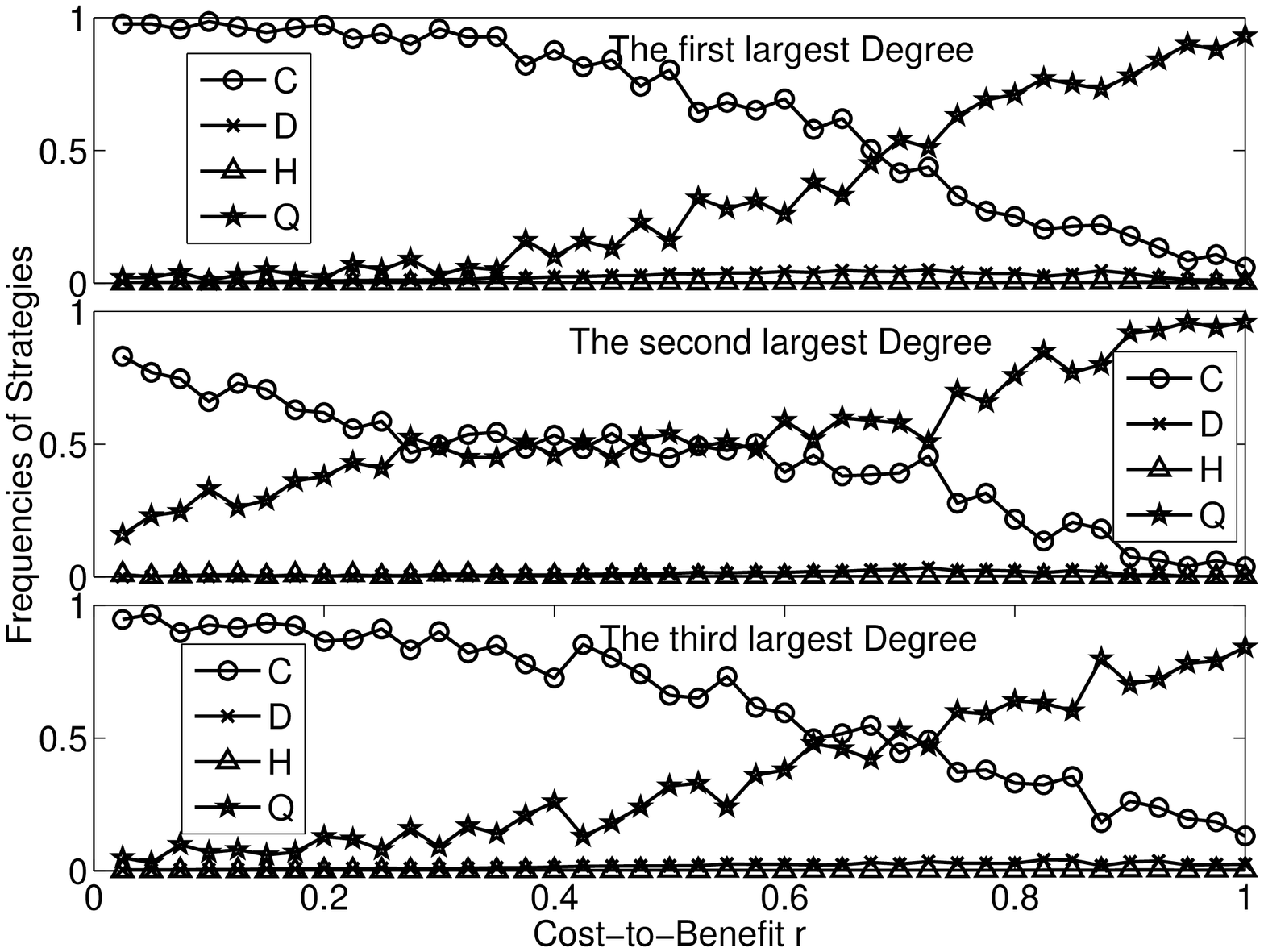}}
\subfigure[]{ \includegraphics[width=0.32\textwidth]{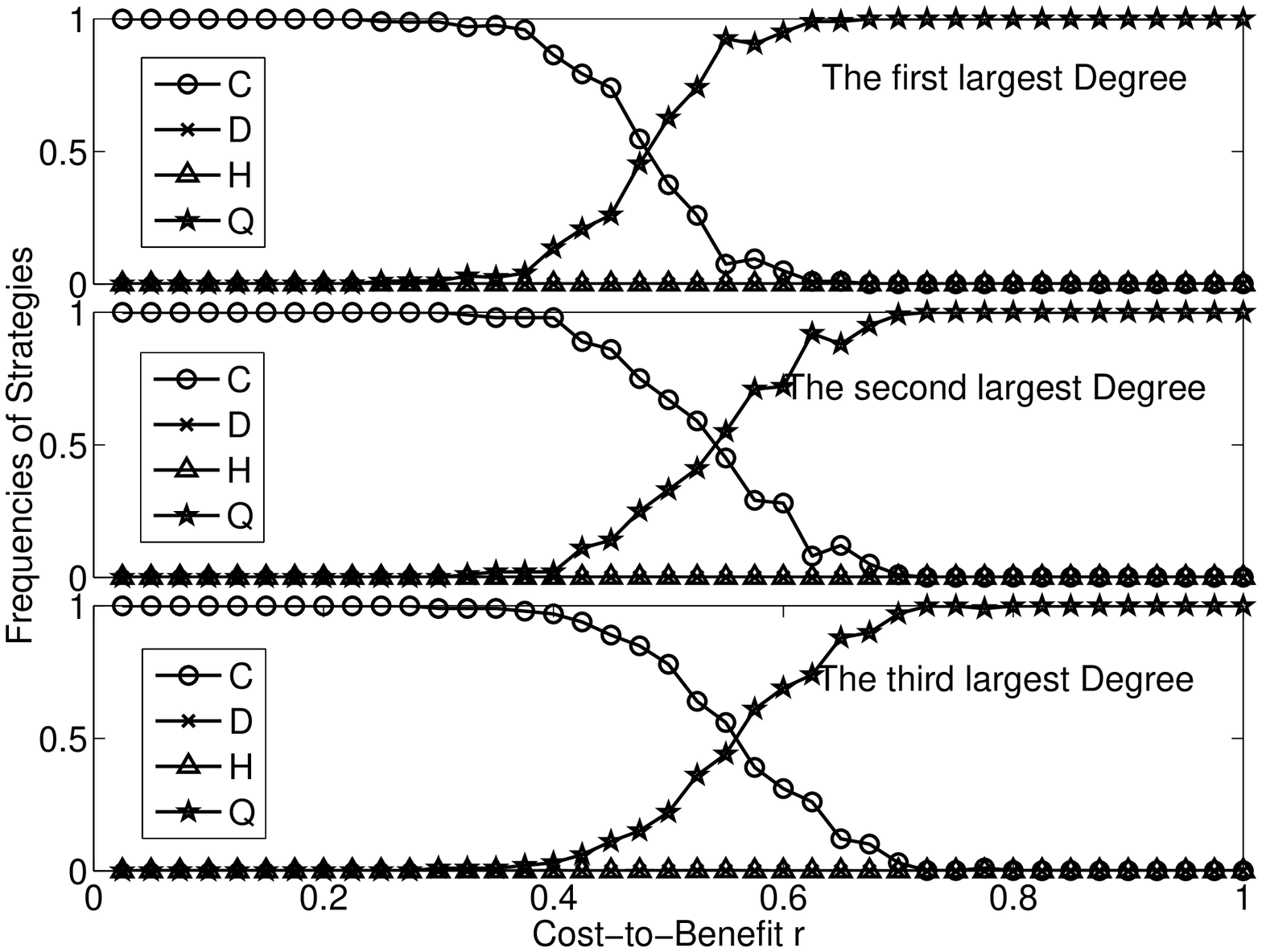}}\\
\caption{Fractions of agents with different strategies in Case 2 on
a SF network, when the agents with the first three largest degrees
play the quantum strategy $\hat{Q}$. (a) Prisoners' Dilemma. (b)
Snowdrift. (c) Stag-Hunt. In the upper, median and lower sub-figures
of (a),(b) and (c), the agent with the first, second or third degree
is assigned the quantum strategy $\hat{Q}$
respectively.}\label{fig4}
\end{figure}

On the other hand, the evolution of strategies on the SF network is
more complex due to the features of the SF network, i.e., the
power-law distribution of degrees of nodes in the network. In both
Case 1 and 2, all strategies fluctuate no matter which game is
played. However, it can be observed that the classical strategies
decrease in proportion with the increase of the variable $T$ or $r$,
while the quantum strategies act conversely. According to the
features of the SF network, if few quantum strategies are played by
some agents with small degrees, it will be harder for them to invade
the network. This raises the question that if a quantum strategy is
employed by a certain agent with one of the first three largest
degrees, namely a hub node, how will the quantum strategies evolve?
In the next simulations, the agent with the first largest degree
will be compulsorily assigned a strategy $\hat{H}$ in Case 1 or
$\hat{Q}$ in Case 2 after each agent chooses a strategy. The results
are shown in the uppermost sub-figure in Fig.~\ref{fig3} and
Fig.~\ref{fig4}, while the other two sub-figures in Fig.~\ref{fig3}
and Fig.~\ref{fig4} are the results when the agent with the second
or the third largest degree plays a strategy $\hat{H}$ or $\hat{Q}$.
This procedure is applied on the PD, SD and SH games, respectively.

From Fig.~\ref{fig3} and Fig.~\ref{fig4}, it can be seen that when
the agent with the first largest degree plays a quantum strategy,
the fluctuations in the results reduce significantly, which means
that if the node with the largest degree is occupied by an agent
with a quantum strategy, then the strategy spreads out more quickly
on the SF network and the population is invaded more easily by the
strategy. If the quantum strategy is assigned to an agent with the
second or the third largest degree, the fluctuations also decrease,
but it is not lower than that of the first one. As for the three
games, the SH game makes the fluctuations of results lower than
those on the PD and SD games regardless of degrees. Furthermore,
comparing the two strategies $\hat{H}$ and $\hat{Q}$, we can find
that the value of the temptation $T$ or the cost-to-benefit $r$ when
the strategy $\hat{Q}$ becomes dominated is smaller than that of the
strategy $\hat{H}$ according to Fig.~\ref{fig3} and Fig.~\ref{fig4}.

\begin{figure}[th]
\centering \subfigure[]{
\includegraphics[width=0.32\textwidth]{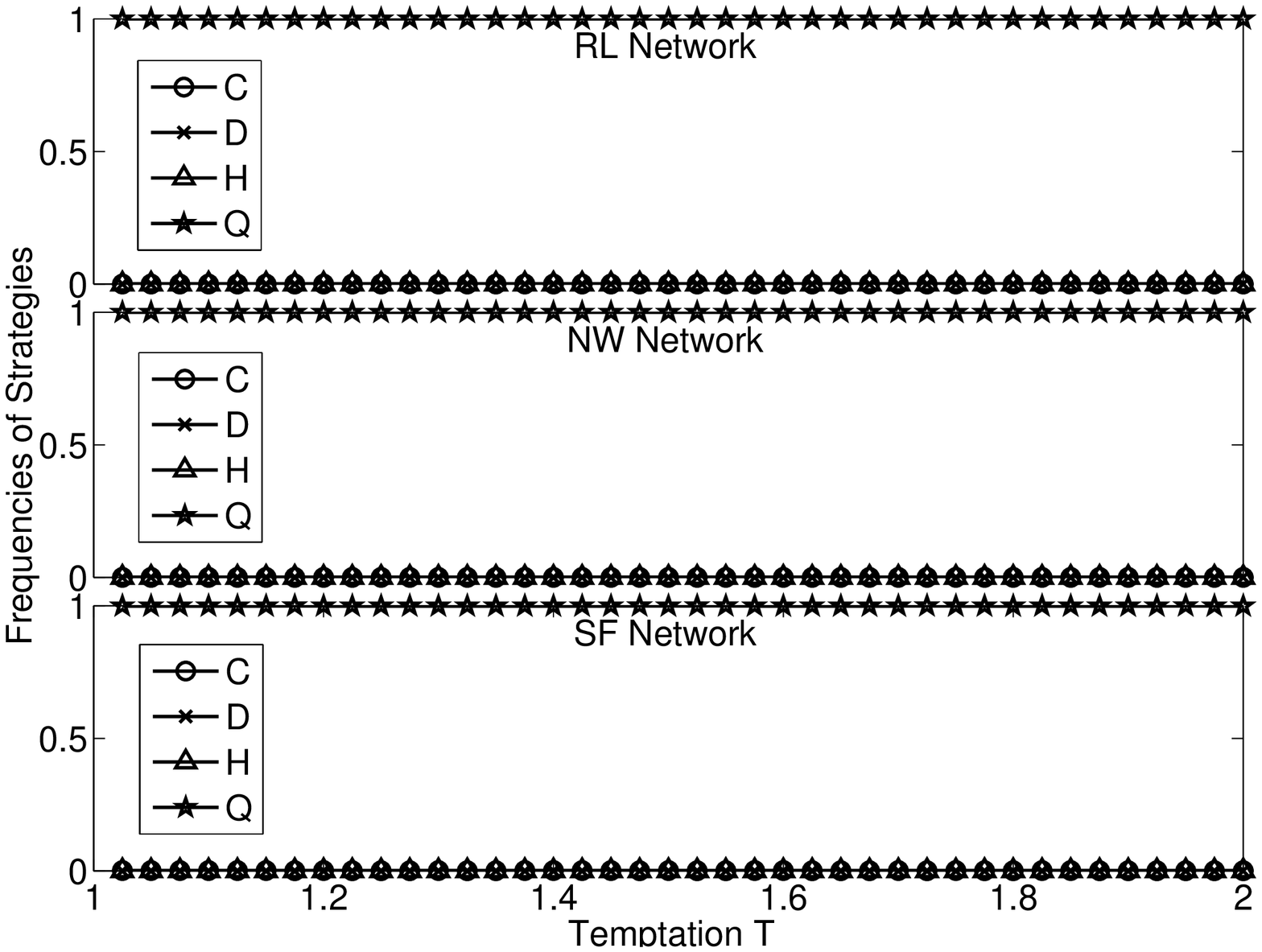}} \subfigure[]{
\includegraphics[width=0.32\textwidth]{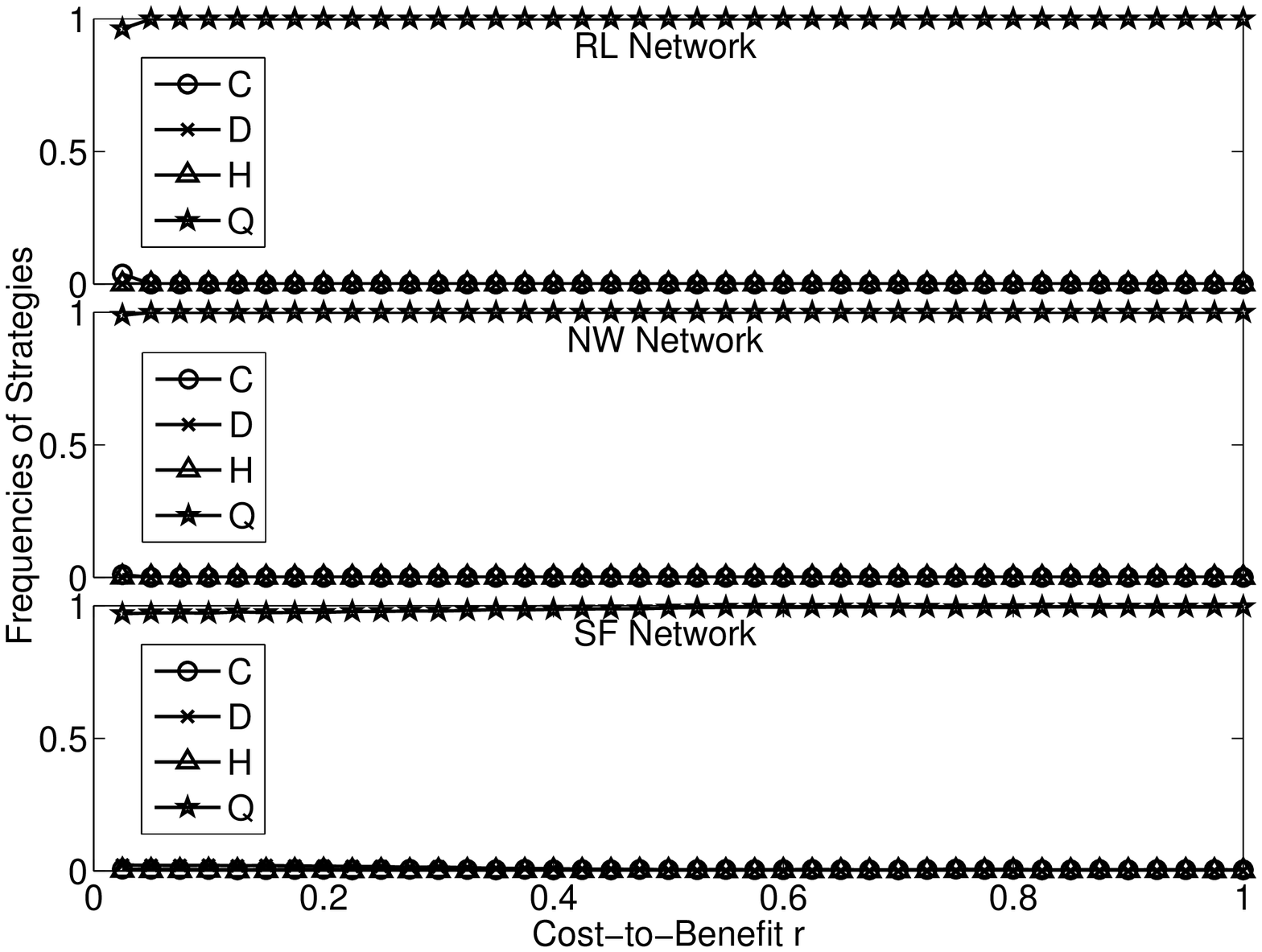}}
\subfigure[]{ \includegraphics[width=0.32\textwidth]{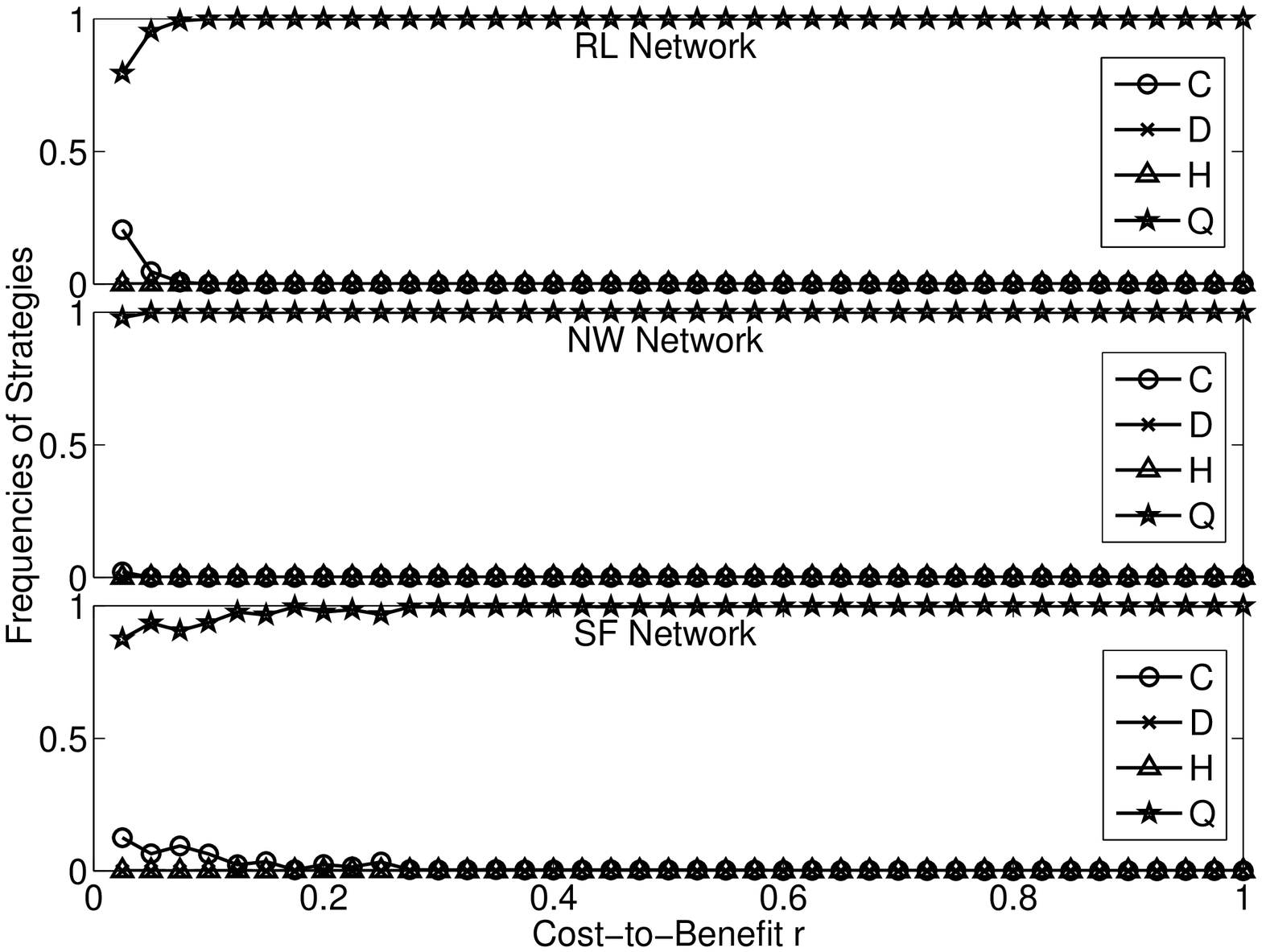}}\\
\caption{Fractions of agents with different strategies in Case 2 on
three games and three networks, when the fraction of the quantum
strategy $\hat{Q}$ is increased at 25\%. (a) Prisoners' Dilemma. (b)
Snowdrift. (c) Stag-Hunt. In the upper, median and lower sub-figures
of (a),(b) and (c), games are played on RL, NW and SF networks
respectively.}\label{fig5}
\end{figure}

For Case 2, the main reason why quantum strategies do not invade the
population from the outset is that initially the fractions of agents
with quantum strategies are too low. If the fraction is increased,
the quantum strategy $\hat{Q}$ will be able to dominate the network.
In simulations, the fractions of the strategy $\hat{D}$ and
$\hat{H}$ remain constant at 49\% and 1\%, while the fractions of
the other two strategies are adjusted. Further, the fraction of the
quantum strategy $\hat{Q}$ is set at 10\%, 20\% and 25\%
respectively and correspondingly the fraction of the strategy
$\hat{C}$ is 40\%, 30\% and 25\%. For the SF network, the agent
occupying the node with the largest degree is assigned a quantum
strategy. Fig.~\ref{fig5} exhibits the evolution of strategies when
the fraction of the quantum strategy $\hat{Q}$ is 25\%, where we can
see that the quantum strategy $\hat{Q}$ can invade the population
successfully on all networks and becomes the ESS from the outset.

\begin{figure}[th]
\centering \subfigure[]{
\includegraphics[width=0.32\textwidth]{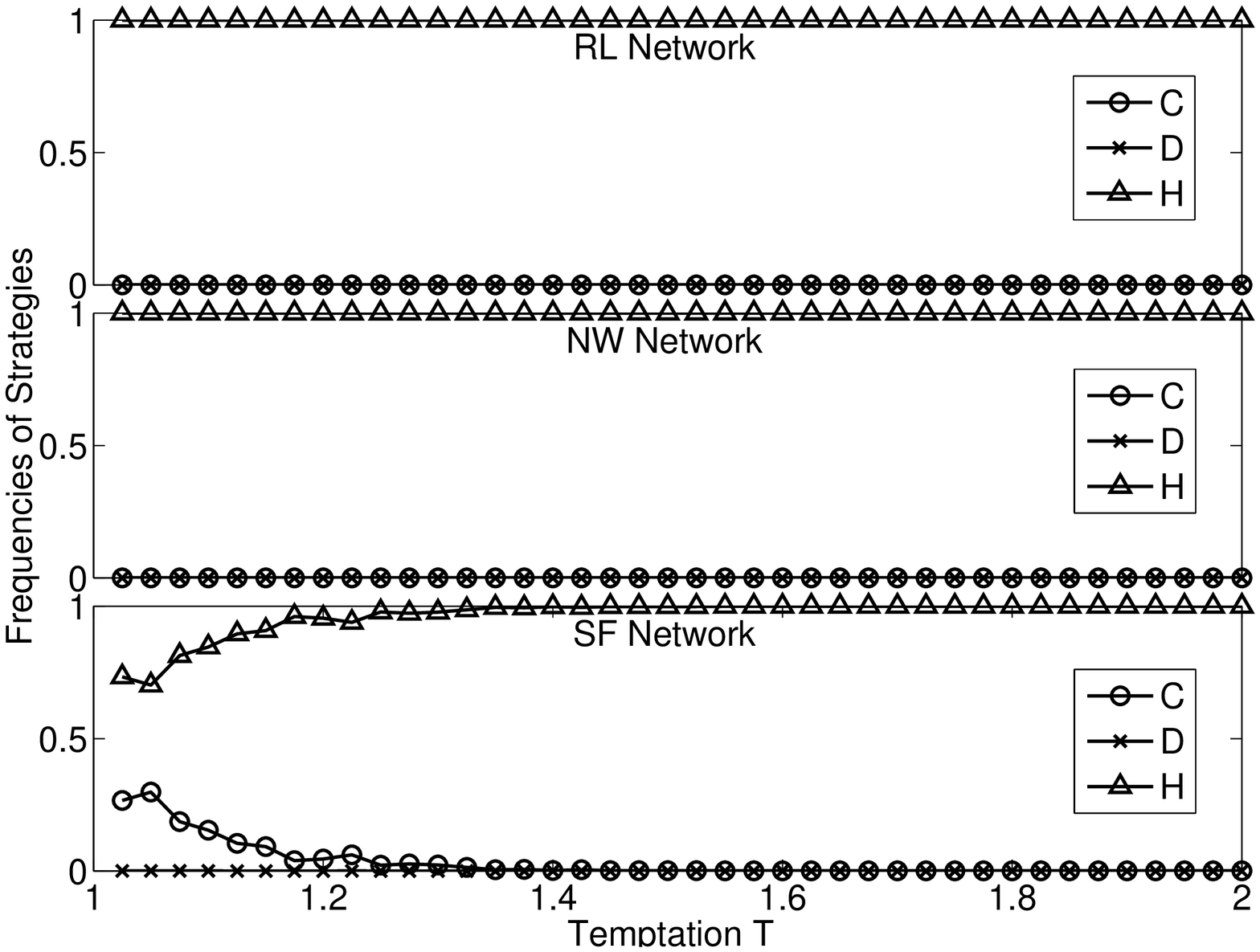}} \subfigure[]{
\includegraphics[width=0.32\textwidth]{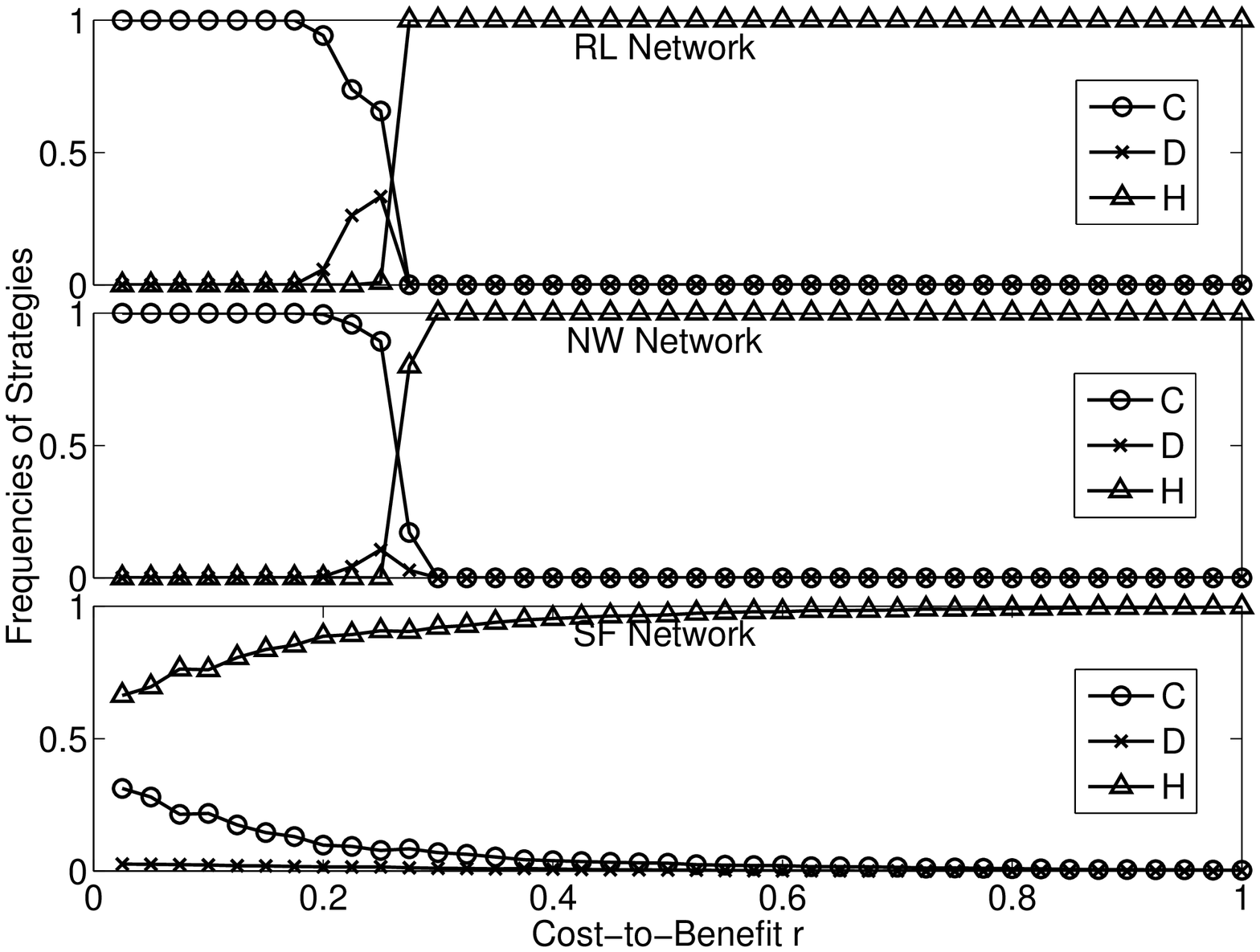}}
\subfigure[]{ \includegraphics[width=0.32\textwidth]{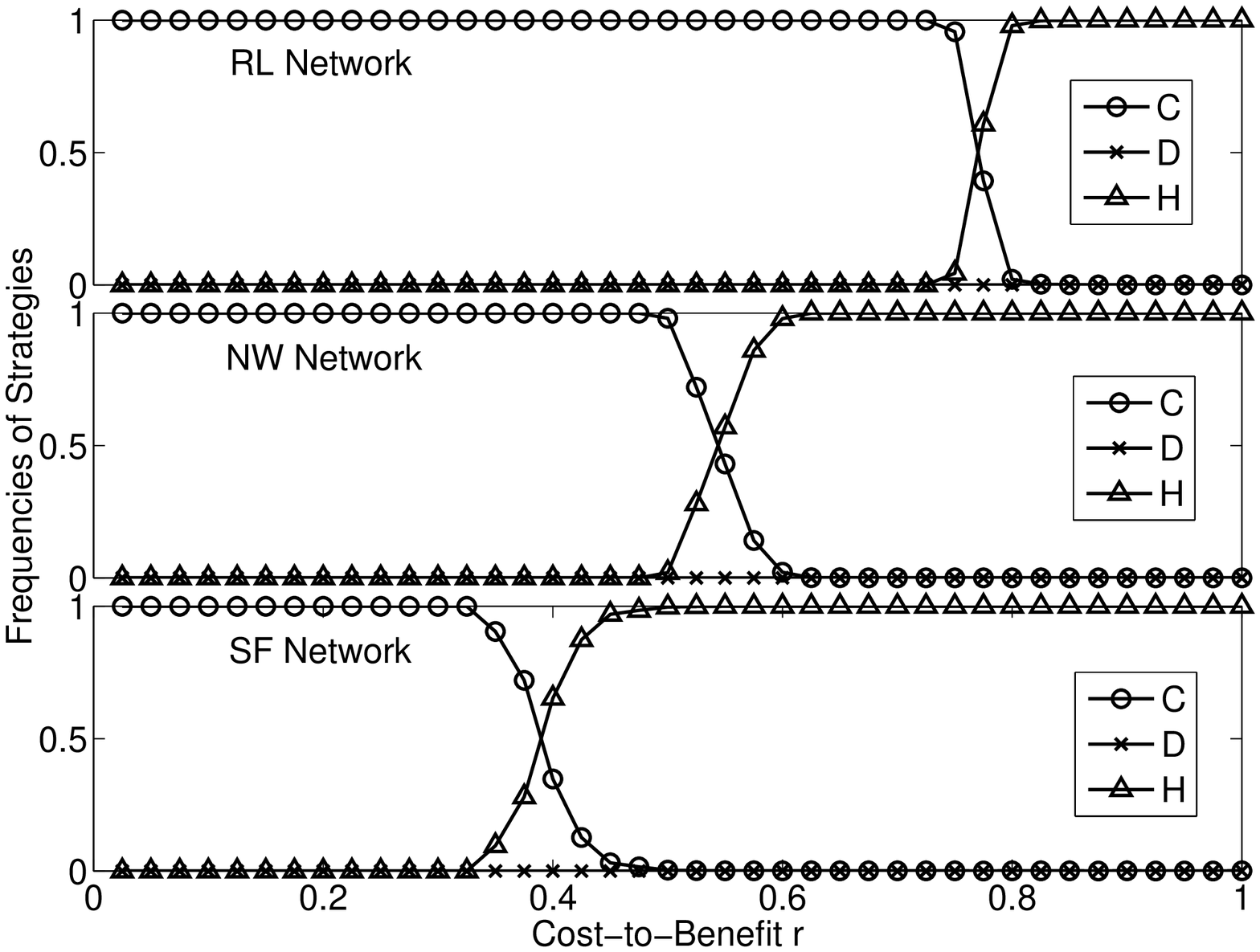}}\\
\caption{Fractions of agents with different strategies in Case 1 on
three games and three networks, when the fraction of the quantum
strategy $\hat{H}$ is increased at 25\%. (a) Prisoners' Dilemma. (b)
Snowdrift. (c) Stag-Hunt. In the upper, median and lower sub-figures
of (a),(b) and (c), games are played on RL, NW and SF networks
respectively.}\label{fig6}
\end{figure}

However, for Case 1, the situation is more complex than that in Case
2. Besides the reason mentioned above, another major reason also
prevents the quantum strategy $\hat{H}$ being spread on networks
that is the strategy profile $(\hat{H},\hat{H})$ is not Pareto
optimal although it is a Nash equilibrium. Hence, even if the
fraction of the quantum strategy $\hat{H}$ is increased to 25\%, it
cannot dominate on all networks from the outset, especially on the
SF network, as is shown in Fig.~\ref{fig6}, in which similarly the
fraction of the quantum strategy $\hat{H}$ is set at 25\% and
correspondingly the fraction of the strategy $\hat{C}$ is also 25\%,
while the fraction of the strategy $\hat{D}$ remains constantly at
50\%.

\section{Conclusions}
In summary, we investigate the evolution of strategies on networks
when quantum strategies $\hat{H}$ and $\hat{Q}$ are employed as
invaders. For the evolution of strategies, the structure of a
network is a decisive factor and a game represents an agent's
response to some external stimuli. So, we construct three networks
and introduce three games in this paper to investigate the evolution
of strategies on these networks in a defector dominated population
when different games are employed. As far as two quantum strategies
are concerned, the strategy $\hat{Q}$ is more aggressive than the
other one regardless of Case 1 or Case 2, because it is not only a
Nash equilibrium but also Pareto optimal. Considering three
networks, we find that the population on a RL network can be invaded
most easily by quantum strategies without any small world effects
(properties), namely short average path length and large clustering
coefficient. On the contrary, in the SF network the power-law
distribution of degrees makes the spread of quantum strategies more
difficult and exacerbates the fluctuations of results when few
quantum strategies happen to be played by some agents only with
small degrees. If an agent with a quantum strategy occupies a hub,
i.e. a node with the largest degree, the fluctuations reduce
considerably. Furthermore, if the fractions of quantum strategies
are increased significantly, they can dominate a network from the
outset.

\section*{Acknowledgments}
This work is supported by the National Natural Science Foundation of
China (Grant No. 61105125 \& No. 51177177) and by the Australian
Research Council (Grant DP0771453).


\bibliographystyle{model1a-num-names}
\bibliography{paper1}







\end{document}